\documentclass[11pt]{article}
\usepackage[latin1]{inputenc}
\usepackage{graphics}
\usepackage{epsfig}

\newcommand{\real}{{I\!\!R}}

\newcommand{\xn}{{\bf x}}

\newcommand{\Kn}{{\bf K}}
\newcommand{\Xn}{{\bf X}}

\newcommand{\yn}{{\bf y}}

\newcommand{\Hn}{{\bf H}}

\newcommand{\Wn}{{\bf W}}

\newcommand{\betn}{{\mbox{\boldmath $\beta$}}}

\newcommand{\mun}{{\mbox{\boldmath $\mu$}}}

\newcommand{\tetn}{{\mbox{\boldmath $\theta$}}}

\begin{document}
\title{\bf A Poisson Mixed Model with Nonnormal Random Effect Distribution}
\author{\renewcommand{\thefootnote}{\arabic{footnote}}
        {\bf Lizandra C. Fabio }\ and {\bf Gilberto A. Paula\footnotemark[1]} \\
        {\em Instituto de Matem\' atica e Estat\'{\i}stica - USP, Brazil}\\
{\centerline{and}}\\
{\bf M\'ario de Castro}\\
{\em Instituto de Ci\^encias Matem\'aticas e de Computa\c c\~ao - USP, Brazil}
}
\date{}
\maketitle
\baselineskip=15pt
\centerline{\bf Abstract}
\noindent We propose in this paper a random intercept Poisson model
in which the random effect distribution is assumed to follow a generalized
log-gamma (GLG) distribution. We derive the first two moments for the marginal
distribution as well as the intraclass correlation. Even though numerical
integration methods are in general required for deriving the marginal models,
we obtain the multivariate negative binomial model  for a particular parameter
setting of the hierarchical model. An iterative process is derived for
obtaining the maximum likelihood estimates for the parameters in
the multivariate negative binomial model.
Residual analysis are proposed and
two applications with real data are given for illustration.

\vspace{2mm}
\noindent{\tt Key words:} Count data; Generalized log-gamma distribution; Multivariate negative binomial distribution;
Overdispersion; Random-effect models.
\vspace{2mm}

\footnotetext[1]{Address for correspondence: Instituto de Matem\'atica e Estat\'{\i}stica,
USP - Caixa  Postal 66281 (Ag. Cidade de S\~ao Paulo),
05314-970 S\~ao Paulo - SP - Brazil. \\ \hspace{3mm} e-mail: lcfabio@ime.usp.br}

\baselineskip=20pt
\section{Introduction}
The effects of the misspecification of the random effect distribution in
generalized linear mixed models (GLMMs) have been investigated by some authors
recently. For instance, Litière et al. (2008) verified by Monte Carlo studies
that the misspecification of the random effect distribution of the response
variable in random intercept
logistic models may lead to severe bias in the random effect component prediction,
which in many problems may be the main parameter of interest. These same authors have
proposed a family of tests to detect the misspecification of the random effect
distribution in GLMMs (Alonso et al., 2008). In addition, Lee and Nelder (1996, 2001)
have suggested a flexibilization of the random effect distribution in GLMMs, but under a
hierarchical framework. Although any
combination between the conditional response and the random effect distributions
may be considered in
the Lee and Nelder's proposal, the majority of the applications have been done
for conjugate distributions. In particular,
under the marginal framework, Molenberghs et al. (2007)
have presented a combination between gamma and normal random effects in Poisson
mixed models and more recently Zhang et al. (2008) assumed a log-gamma distribution for the
random effects in linear mixed models.

The aim of this paper is to present an alternative distribution for
the random effect in random intercept Poisson models, which is
characterized by assuming a generalized log-gamma distribution for the
random effect component. This distribution
introduced by Prentice (1974) (see also Lawless, 1980)
 has as particular cases the normal and
extreme value distributions and
it may assume skew forms to the right and to the left. In addition, generalized log-gamma models
have been widely applied in the areas
of survival analysis and reliability.
For instance, DiCiccio (1987) derived approximate inferences for the quantiles
and scale parameters whereas Young and Bakir (1987) obtained the bias of order $n^{-1}$, where
$n$ is the sample size, for the parameter estimates in generalized log-gamma
regression models for uncensored samples. Young and Bakir (1987) also presented the expectation of various log-likelihood derivatives in closed-form expressions. Ahn (1996) proposed
a regression tree method to classify the heterogeneous subsets of the data into different
generalized log-gamma regression models with the shape parameter being estimated separately in each formed stratum under independent random censoring.
Ortega et al. (2003) derived the appropriate matrices for
assessing local influence on the parameter estimates under
different perturbation schemes and  Chien-Tai et al. (2004)
presented a conditional method of inference to derive confidence intervals for the location
as well as quantiles and reliability functions under progressively type-II censoring and
by assuming the shape parameter known in generalized log-gamma regression models
with censored data.
More recently, Cox et al. (2007) presented a taxonomy of the hazard function
of generalized gamma distribution with application to study of survival after diagnosis of
clinical AIDS during different phases of HIV therapy and Ortega et al. (2009)
introduced the generalized log-gamma regression models with cure fraction giving
emphasis to assessment of local influence.

The paper is organized as follows. In Section 2 we present a brief review of the
generalized log-gamma distribution. The random intercept Poisson generalized log-gamma
model is proposed in Section 3, as well as a discussion on the parameter and random
effect estimation. The derivation of the first two moments for the
marginal distribution and of the intraclass correlation are given in Section 4.
For a particular parameter setting of the hierarchical model we derive,
in Section 5, the multivariate negative
binomial model (Johnson et al., 1997) as a marginal model.
An iterative process for the parameter estimation as well as goodness-of-fit procedures
and residual analysis are also presented in Section 5.
In Section 6 the epilepsy data set
(Diggle et al., 2002) is fitted with the random intercept Poisson-normal
and random intercept Poisson-GLG
models and compared under the AIC criterion. Another application that has been
analyzed by Poisson models (Lange et al., 1994)
is reanalyzed with the negative binomial and multivariate negative binomial models.

\section{Generalized log-gamma distribution}
Let $y$ be a random variable  following a generalized log-gamma distribution. The
probability density function (pdf)  of $y$ (see, for instance, Lawless, 2002) is given by
\begin{eqnarray}
f(y; \mu, \sigma, \lambda)=\left \{ \begin{array}{ll}\frac{c(\lambda)}{\sigma}
{\rm exp}\left [\frac{(y - \mu)}{\lambda \sigma} - \frac{1}{\lambda^2}
{\rm exp} \left \{ \frac{\lambda(y - \mu)}{ \sigma} \right \} \right ],
  & \textrm{if} \quad  \lambda \ne  0\\
\frac{1}{\sigma\sqrt{2\pi}}{\rm exp}\left \{-\frac{(y - \mu)^2}{2\sigma^2}
\right \}, &\textrm{if} \quad
\lambda=0,
\end{array}\right.
\end{eqnarray}
where $y\in \real;$ $\mu \in \real,$ $\sigma>0$ and $\lambda \in \real$ are, respectively, the position, scale and shape parameters
and $c(\lambda) = \frac{|\lambda|}{\Gamma(\lambda^{-2})}
(\lambda^{-2})^{\lambda^{-2}}$ with $\Gamma(\cdot)$ being the gamma function.
We will denote
$y \sim {\rm GLG}(\mu, \sigma, \lambda)$.
The extreme value distribution is a particular
case of (1), when $\lambda=1$. For $\lambda < 0$ the pdf of $y$ is
skew to the right and for $\lambda > 0$ it is skew to the left.
Figure 1 presents the behavior of the pdf of
$y\sim{\rm GLG}(0,1,\lambda)$ for some values of $\lambda$.

One has for $\lambda \ne 0$ the following moments:
$$
    {\rm E}(y) =  \mu + \sigma \left \{ \frac{\psi(\lambda^{-2}) + {\rm log} \lambda^{-2}}
{|\lambda|} \right \}  \ \ \  \  {\rm and} \ \ \ \
{\rm Var}(y) =  \frac{\sigma^2 \psi'(\lambda^{-2})}{\lambda^2},
$$
where $\psi(\cdot)$ and $\psi'(\cdot)$ denote, respectively, the digamma and trigamma
functions. For $\lambda = 0$ one has the normal case for which
${\rm E}(y) =  \mu$ and
${\rm Var}(y) = \sigma^2$.

\vspace{-0.5cm}
\begin{figure}[h]
\centerline{\hbox{\epsfig{file=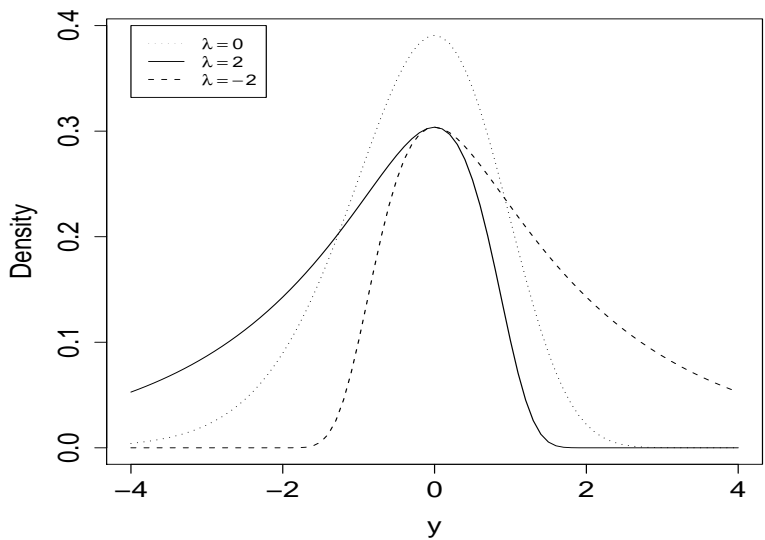,height=3.2in,width=4.2in}}}
\caption[]{Graphs of the generalized log-gamma distribution for some values of $\lambda$ and by assuming
$\mu=0$  and $\sigma = 1$.}
\label{fig41}
\end{figure}

\section{The random intercept Poisson-GLG model}
Let $y_{ij}$ denote the $j$th outcome measured for the $i$th
cluster (subject), $i=1, \ldots, n$ and $j=1, \ldots, m_i$. We will assume
the following random intercept Poisson-GLG model:
\begin{description}
\item (i) $y_{ij} |b_i$  $\stackrel{\rm ind} {\sim}$ ${\rm Poisson}(u_{ij})$
\item (ii) $u_{ij}$ =  ${\rm exp}(\xn_{ij}^{\top}\betn + b_i)$  and
\item (iii) ${b_i} \stackrel{\rm iid} {\sim} {\rm GLG}(0, \sigma, \lambda)$,
\end{description}
where $\xn_{ij} = (x_{ij1}, \ldots, x_{ijp})^{\top}$
contains values of explanatory variables and $\betn = (\beta_1, \ldots, \beta_p)^{\top}$
is the parameter vector of the systematic component. The model (i)-(iii) will be named
random intercept Poisson-GLG model.
When $\lambda=0$ one has the random intercept  Poisson-normal model
(see, for instance, Breslow and Clayton, 1993).
Let $f_{Y|b}(y_{ij}|{ b_i}, \betn)$ and $f_b({ b_i};\sigma, \lambda)$ be the probability function
 of $y_{ij}|{ b_i}$
and the pdf of ${b_i}$, respectively. Then, the marginal probability function of
$\yn = (\yn_1, \ldots, \yn_n)^{\top}$, where
$\yn_i = (y_{i1}, \ldots, y_{im_i})^{\top}$,
is given by
$$
  f_Y(\yn ; \betn, \sigma, \lambda) =
  \prod_{i=1}^n f_Y(\yn_i ; \betn, \sigma, \lambda)
$$
with
\begin{equation}
  f_Y(\yn_i ; \betn, \sigma, \lambda) =
  \int_{-\infty}^{+\infty} \left\{\prod_{j=1}^{m_i} f_{Y|b}(y_{ij}|{b_i}, \betn)\right\}f_b({b_i};\sigma, \lambda)d{ b_i},
\end{equation}
which in general does not have a closed-form. Then, the log-likelihood function for the
marginal model, using (2), takes the form
\begin{equation}
{\rm L}(\tetn) = \sum_{i=1}^n {\rm log}\int_{-\infty}^{+\infty}\left\{\prod_{j=1}^{m_i} f_{Y|b}(y_{ij}|{ b_i}, \betn)\right\}f_b({b_i};\sigma, \lambda)d{b_i},
\end{equation}
where $\tetn = (\betn^{\top}, \lambda, \sigma)^{\top}$. Expression (3) should be
approximated by numerical integration methods, such as Laplace approximation
or Gauss-Hermite quadrature. To predict the random effects we can use the empirical Bayes method
(see, for instance, McCulloch and Searle, 2001) given by
$$
\tilde {b_i} = {\rm E} [ { b_i} | \yn_i ]
= \frac{\int_{-\infty}^{+\infty} { b_i} f_{Y|b}(\yn_i | { b_i}, \betn) f_b({ b_i};\sigma, \lambda) d{b_i}}
{\int_{-\infty}^{+\infty} f_{Y|b}(\yn_i | {b_i}, \betn) f_b({ b_i};\sigma, \lambda) d{ b_i}}.
$$
The NLMIXED procedure available in SAS has been required to obtain the parameter estimate in the GLMM class. Through this procedure it is possible to compute the integral in (3) and to perform the maximization of the log likelihood in (3) as well as to obtain the random effect prediction.

\section{Derivation of moments}
We derive in this section the moments E$(y_{ij})$ and Var$(y_{ij})$ as well as
Cov$(y_{ij}, y_{ij'})$, for $j \ne j'$, and the following results will be used:
\begin{description}
\item (a)  E$(y_{ij})$ = E$\{$E$(y_{ij} | { b_i})\}$ = $\mu_{ij}$E$(e^{{ b_i}})$,
\item (b) Var$(y_{ij})$ = Var$\{$E$(y_{ij} | { b_i})\}$ + E$\{$Var$(y_{ij} | { b_i})\}$ =
$\mu_{ij}^2$Var$(e^{{ b_i}})$ + $\mu_{ij}$E$(e^{{ b_i}})$  and
\item (c) Cov$(y_{ij}, y_{ij'})$ = Cov$\{$E$(y_{ij}|{ b_i})$,E$(y_{ij'}|{ b_i})\}$ +
E$\{$Cov$(y_{ij}, y_{ij'} | {b_i})\}$ =
\\ = Cov$(\mu_{ij}e^{{ b_i}}, \mu_{ij'}e^{{ b_i}})$ + 0 = $\mu_{ij} \mu_{ij'}$
Var$(e^{{ b_i}})$, for $j \ne j'$,
\end{description}
where $\mu_{ij} = {\rm exp}(\xn_{ij}^{\top}\betn)$.
From (a)-(c) above one has that
$$
\frac {{\rm Var}(y_{ij})}{{\rm E}(y_{ij})} = 1 +
\mu_{ij}\frac {{\rm Var}(e^{{ b_i}})}{{\rm E}(e^{{ b_i}})}
$$
and since
$\mu_{ij} > 0$, Var$(e^{{ b_i}})>0$ and E$(e^{{ b_i}})>0$ it follows
that Var$(y_{ij})$ $>$ E$(y_{ij})$, that is, the model (i)-(iii) is overdispersed.
In Table 1 one has the expressions of E$(e^{{ b_i}})$ and E$(e^{2{ b_i}})$ for
some ranges of $\lambda$, where
$$
{\rm I}_1(\lambda, \sigma) = \int_0^{\infty} t^{\lambda^{-2}(\lambda \sigma + 1) - 1}e^{-t}dt
\ \ {\rm and} \  \
{\rm I}_2(\lambda, \sigma) = \int_0^{\infty} t^{\lambda^{-2}(2\lambda \sigma + 1) - 1}e^{-t}dt,
$$
which should be solved by numerical integration methods.
\begin{center}
\begin{tabular}{ccc}\\
\multicolumn{3}{c}{\bf Table 1}\\
\multicolumn{3}{c}{First two moments for the
random variable $e^{{ b_i}}$ according to the values of the}\\
\multicolumn{3}{c}{shape parameter $\lambda$.}\\ \hline
Shape parameter & E$(e^{{ b_i}})$ & E$(e^{2{ b_i}})$ \\ \hline
    $\lambda=0$   &  $e^{\sigma^2/2}$ &  $e^{2\sigma^2}$ \\
    $\lambda>0$   &  $\frac{(\lambda^2)^{\sigma/\lambda}}{\Gamma(\lambda^{-2})}
\Gamma\{\lambda^{-2}(\lambda \sigma + 1) \}$    &
$\frac{(\lambda^2)^{2\sigma/\lambda}}{\Gamma(\lambda^{-2})}\Gamma\{\lambda^{-2}(2\lambda \sigma + 1)\}$ \\
$\lambda<0$   &  $\frac{(\lambda^2)^{\sigma/\lambda}}{\Gamma(\lambda^{-2})}
{\rm I}_1(\lambda, \sigma)$ &
$\frac{(\lambda^2)^{2\sigma/\lambda}}{\Gamma(\lambda^{-2})}{\rm I}_2(\lambda, \sigma)$
                \\ \hline
\end{tabular}
\end{center}

\section{The multivariate negative binomial model}
Consider now the following random intercept Poisson-GLG
model:
\begin{description}
\item (i) $y_{ij} | {b_i}$  $\stackrel{\rm ind} {\sim}$ ${\rm Poisson}(u_{ij})$,
\item (ii) $u_{ij}$ =  ${\rm exp}(\xn_{ij}^{\top}\betn + {b_i})$   and
\item (iii) ${b_i} \stackrel{\rm iid} {\sim} {\rm GLG}(0, \lambda, \lambda),$
\end{description}
that is, the same model given in Section 3 with
$\sigma = \lambda$ $(\lambda > 0)$.

Denoting $\phi = \lambda^{-2}$ the joint distribution of $(\yn_i, { b_i})$ is given by
\begin{eqnarray*}
  f_Y(\yn_i, { b_i}; \betn, \phi)& = &  \left\{\prod_{j=1}^{m_i} f_{Y|b}(y_{ij}|{ b_i}, \betn)\right\}
f_b({ b_i};\phi) \\
&=& \frac { \phi^{\phi} (\prod_{j=1}^{m_i} \mu_{ij}^{y_{ij}})}
{\Gamma(\phi) (\prod_{j=1}^{m_i} y_{ij}!)}
{\rm exp}\{- \mu_{i+}{\rm exp}({ b_i}) - \phi{\rm exp}({ b_i}) + { b_i}y_{i+} + { b_i} \phi\},
\end{eqnarray*}
where $y_{i+} = \sum_{j=1}^{m_i} y_{ij}$ and
$\mu_{i+} = \sum_{j=1}^{m_i} \mu_{ij}$.
Consider the variable transformation $t_i = {\rm exp}({ b_i})(\mu_{i+} + \phi)$. One
has that $\frac{d{ b_i}}{dt_i} = \frac{1}{t_i}$ and
the joint distribution of $(\yn_i, t_i)$ assumes the form
\begin{eqnarray*}
  f(\yn_i, t_i; \betn, \phi)
&=& \frac {\phi^{\phi} (\prod_{j=1}^{m_i} \mu_{ij}^{y_{ij}})}{\Gamma(\phi) (\prod_{j=1}^{m_i} y_{ij}!)} e^{-t_i} \left( \frac{t_i}{\mu_{i+}
+ \phi} \right)^{y_{i+} + \phi} \frac{1}{t_i} \\
&=& \frac {\phi^{\phi} (\prod_{j=1}^{m_i} \mu_{ij}^{y_{ij}})}
{\Gamma(\phi) (\prod_{j=1}^{m_i} y_{ij}!)(\phi + \mu_{i+})^{(\phi + y_{i+})}}
e^{-t_i}t_i^{y_{i+} + \phi - 1}.
\end{eqnarray*}
Thus, the marginal probability function of $\yn_i$ yields
\begin{eqnarray*}
  f_Y(\yn_i; \betn, \phi)
&=& \frac {\phi^{\phi} (\prod_{j=1}^{m_i} \mu_{ij}^{y_{ij}})}
{\Gamma(\phi) (\prod_{j=1}^{m_i} y_{ij}!)(\phi + \mu_{i+})^{(\phi + y_{i+})}}
\int_0^{\infty} e^{-t_i}t_i^{y_{i+} + \phi - 1}dt_i
\end{eqnarray*}
and since $\Gamma(\phi + y_{i+}) = \int_0^{\infty} e^{-t_i}t_i^{y_{i+} + \phi - 1}dt_i$,
the marginal probability function of $\yn_i$ reduces to
\begin{equation}
f_Y({\bf y}_i ;\betn , \phi) =
\frac{\Gamma(\phi + y_{i+}) \phi^{\phi}}{(\prod_{j=1}^{m_i} y_{ij}!) \Gamma(\phi)}
\frac{{\rm exp}(\sum_{j=1}^{m_i} y_{ij}
{\rm log}\mu_{ij})}{(\phi + \mu_{i+})^{(\phi + y_{i+})}},
\end{equation}
that is the multivariate negative binomial distribution (see, for instance,
Johnson et al., 1997) of means
E$(y_{ij}) = \mu_{ij},$ variances
Var$(y_{ij}) = \mu_{ij} + \frac{\mu_{ij}^2}{\phi}$,
for $j=1, \ldots, m_i$, and covariances
Cov$(y_{ij}, y_{ij'}) = \frac{\mu_{ij} \mu_{ij'}}{\phi}$, for $j \ne j'$.
The intraclass correlation between $y_{ij}$ and $y_{ij'}$, for $j \ne j'$,
can be expressed as
$$
{\rm Corr}(y_{ij}, y_{ij'}) = \frac{\sqrt{\mu_{ij} \mu_{ij'}}}{
\sqrt{(\phi + \mu_{ij})} \sqrt{(\phi + \mu_{ij'})}}.
$$
These correlations are always positive and for large values of $\phi$ the multivariate negative
binomial counts $y_{ij}'s$ behave approximately as independent Poisson observations with respective
means $\mu_{ij}'s$.

Therefore, we derive the multivariate negative binomial distribution from an alternative
way, by assuming a particular log-gamma distribution for the random effect in a random intercept
Poisson model.
By a similar calculation one may show that the marginal distribution of $y_{ij}$ is
a negative binomial distribution of mean $\mu_{ij}$, variance
$\mu_{ij} + \frac{\mu_{ij}^2}{\phi}$ and dispersion parameter $\phi > 0$ (see, for
instance, McCullagh and Nelder, 1989). We will denote
$\yn_i \sim {\rm MNB}(\mun_i, \phi)$, where
$\yn_i = (y_{i1}, \ldots, y_{im_i})^{\top}$,
$\mun_i = (\mu_{i1}, \ldots, \mu_{im_i})^{\top}$ and $\phi > 0$. The multivariate negative
binomial model is defined by assuming  that log$ \mu_{ij} = \xn_{ij}^{\top}\betn$.
Then, the log-likelihood function for the multivariate negative binomial model
yields
\begin{eqnarray}
{\rm L}(\tetn)& = & \sum_{i=1}^n {\log}\left \{ \frac{\Gamma(\phi + y_{i+})}
{\Gamma(\phi)} \right \} - \sum_{i=1}^n \sum_{j=1}^{m_i}{\rm log}y_{ij}!
+ n\phi {\rm log}\phi
- \phi \sum_{i=1}^n {\rm log}(\phi + \sum_{j=1}^{m_i} e^{\xn_{ij}^{\top} \betn}) \nonumber \\
 && + \sum_{i=1}^n \sum_{j=1}^{m_i}y_{ij} {\rm log}
\left ( \frac{e^{\xn_{ij}^{\top} \betn}}{\phi + \sum_{j=1}^{m_i}
e^{\xn_{ij}^{\top} \betn}} \right),
\end{eqnarray}
where $\tetn = (\betn^{\top}, \phi)^{\top}$.
The score function and the Fisher
information matrix may be obtained for $\tetn$ and
an iterative process can be performed to get the maximum likelihood estimates.

\subsection{Score function}
The score function is obtained from (5) by
derivating the log-likelihood function L$(\tetn)$ with respect to $\betn$ and
$\phi$, respectively. We obtain
\begin{eqnarray*}
{\bf U}_{\beta} &=& \sum^{n}_{i=1}\left\{\sum^{m_i}_{j=1}y_{ij}x_{ij} - \frac{(\phi + y_{i+})}{(\phi + \mu_{i+})}\sum^{m_i}_{j=1}x_{ij}\mu_{ij}\right\} \\
   &=&\sum^{n}_{i=1}\Xn^{T}_i(\yn_i - a_i\mun_i) \ \ \ {\rm and}
\end{eqnarray*}
by using the result $\Gamma(\phi + y_{i+})/\Gamma(\phi) =
\phi(\phi + 1)(\phi +2 ) \ldots (\phi + y_{i+} - 1)$
(see, for instance, Lawless, 1987) we find
$$
     {\rm U}_{\phi} = \sum^{n}_{i = 1}\left\{\sum^{(y_{i+}-1)}_{j=0} (j + \phi)^{-1}
- \frac{y_{i+}}{(\phi + \mu_{i+})} - {\rm log}(1 + \phi^{-1}\mu_{i+}) +  \frac{\mu_{i+}}{(\phi + \mu_{i+})} \right\},
$$
where $a_i = \frac{(1 + \phi^{-1}y_{i+})}{(1 + \phi^{-1}\mu_{i+})}, $ ${\bf X}_i$ is an $m_i\times p$ matrix
of rows $\xn_{ij}^{\top}$, for $j=1, \ldots, m_i$, so that
$\sum^{(y_{i+}-1)}_{j=0}$ is zero when $y_{i+} - 1 <0.$

\subsection{Fisher information matrix}
The Fisher information matrix for $\betn$ is obtained such that
\begin{eqnarray*}
     \Kn_{\beta \beta} & = & {\rm E} \left \{ - \frac{\partial {\bf U}_{\beta}}{ \partial \betn^T }
\right \} \\
&=& {\rm E}\left\{\sum^{n}_{i = 1}
\left[\frac{(\phi + y_{i+})}{(\phi + \mu_{i+})}\sum^{m_i}_{j = 1}x_{ijk}x_{ijl}\mu_{ij}
-\frac{(\phi + y_{i+})}{(\phi + \mu_{i+})^2}\sum^{m_i}_{j = 1}x_{ijk}\mu_{ij}\sum^{m_i}_{j = 1}x_{ijl}\mu_{ij}\right] \right\} \\
      & = & \sum_{i=1}^n{\bf X}^{\top}_i\left\{{\bf D}(\mun_i) - (\phi + \mu_{i+})^{-1}\mun_i\mun^{T}_i\right\}{\bf X}_i,
\end{eqnarray*}
in that, ${\bf D}(\mun_i) = {\rm diag}\left\{\mu_{i1}, \ldots, \mu_{im_i}\right\}.$
The calculation of the Fisher information for $\phi$ follows the similar steps of the negative binomial model
(see, for instance, Lawless, 1987). We find
\begin{eqnarray*}
    {\rm K}_{\phi \phi}& = &{\rm E} \left \{ - \frac{\partial {\rm U}_{\phi}}{ \partial \phi }\right\} \\
     & = &{\rm E}\left\{\sum^{n}_{i = 1}\left[\sum^{(y_{i+}-1)}_{j = 0}(j + \phi)^{-2} - \frac{y_{i+}}{(\phi + \mu_{i+})^2} - \frac{\phi^{-1}\mu_{i+}}{(\phi + \mu_{i+})} + \frac{\mu_{i+}}{(\phi + \mu_{i+})^2}\right]\right\} \\
     & = &\sum^{n}_{i = 0}\left\{\sum^{\infty}_{j = 0}(j + \phi)^{-2}{\rm P}(Y_{i+}\geq j) - \frac{\phi^{-1}\mu_{i+}}{\mu_{i+} + \phi}\right\}.
\end{eqnarray*}
In addition, it may be showed the orthogonality between $\betn$ and $\phi$, as in the univariate case.
Thus, the Fisher information matrix for $\tetn$ takes the block-diagonal form
$\Kn_{\theta \theta} = {\rm diag}\{\Kn_{\beta \beta}, {\rm K}_{\phi \phi} \}$.

\subsection{Iterative process}
Similarly to the univariate case we can perform a scoring Fisher and a Newton-Raphson iterative processes for
obtaining the maximum likelihood estimates $\hat \betn$ and $\hat \phi$, respectively, which are given by
\begin{equation} \label{eq6}
   \betn^{(r+1)} = \betn^{(r)} + \left\{\sum_{i=1}^{n}\Xn^{\top}_i\Wn^{(r)}_i\Xn_i\right\}^{-1}\sum_{i=1}^{n}\Xn^{\top}_i
\left\{ \yn_i - \frac{(\phi^{(r)} + y_{i+})}{(\phi^{(r)} + \mu^{(r)}_{i+})}\mun^{(r)}_i\right\}
\end{equation}
and
\begin{equation} \label{eq7}
   \phi^{(r+1)} = \phi^{(r)} - \frac{{{\rm U}_{\phi}}^{(r)}}{\ddot {\rm L}_{\phi \phi}^{(r)}}, \ \ {\rm for} \ \ r =0, 1, 2, \ldots.
\end{equation}
in that ${\bf W}_i = {\bf D}(\mun_i) - (\phi + \mu_{i+})^{-1}\mun_i\mun^{T}_i$
and ${\ddot {\rm L}_{\phi \phi}} = \partial {\rm U}_{\phi}/\partial \phi$.
 We can start the iterative process defined in  (\ref{eq6}) and (\ref{eq7})
by using, for instance, the maximum likelihood estimates from the univariate case in which the $m_i$ observations for each group are assumed independent. For large sample ($n$ large) we expected that the maximum likelihood estimators follow,
under suitable regularity conditions, normal distributions. That is, for $n$ large $\hat{\betn} \sim {\rm N}_p(\betn, {\bf K}^{-1}_{\beta \beta})$ and $\hat{\phi} \sim {\rm N}(\phi,{\bf K}^{-1}_{\phi \phi} ).$
In addition, due to the orthogonality between $\betn$ and $\phi$ one has the asymptotic independence
between $\hat \betn$ and $\hat \phi$.

\subsection{Residual analysis}
We found that the estimates of the multivariate negative binomial saturated model log-likelihood function is $\hat{\mu}^{0}_{ij} = y_{ij},$ and therefore, the MNB deviance function has the following expression,
\begin{equation} \label{eq8}
D(\yn, \hat \mun_i, \phi) =
\sum_{i=1}^n2 \left [ \phi{\rm log}\left \{ \frac{\phi + \hat \mu_{i+}}{\phi + y_{i+}} \right \} +
\sum_{j=1}^{m_i}y_{ij} {\rm log} \left \{ \frac{y_{ij}(\phi + \hat \mu_{i+})}{\hat \mu_{ij}(\phi + y_{i+})} \right \} \right ].
\end{equation}
Similarly to the univariate case (see, for instance, Svetliza and Paula, 2003) we can define the
deviance as a measure for multivariate negative binomial models from (\ref{eq8}). So, after some
algebraic manipulation and by assuming that $\phi$ is fixed and $y_{ij} > 0$, $\forall ij$, we
may express the deviance as D$(\yn; \hat \mun) = \sum_{i=1}^n \sum_{j=1}^{m_i} d^2(y_{ij}, \hat \mu_{ij}, \phi)$,
where
\begin{eqnarray} \label{eq9}
d^2(y_{ij}, \hat \mu_{ij}, \phi)=
\left \{ \begin{array}{ll}2 \left [ \frac{\phi}{m_i} {\rm log}\left \{ \frac{\phi + \hat \mu_{i+}}{\phi + y_{i+}} \right \} +
y_{ij} {\rm log} \left \{ \frac{y_{ij}(\phi + \hat \mu_{i+})}{\hat \mu_{ij}(\phi + y_{i+})} \right \} \right ] & \textrm{if} \quad  y_{ij} \ne  0\\
\frac{2\phi}{m_i} {\rm log} \left \{ \frac{\phi + \hat \mu_{i+}}{\phi + y_{i+}} \right \} &\textrm{if} \quad
y_{ij}=0,
\end{array}\right.
\end{eqnarray}
The quantity $\phi$ that appears in the deviance expression may be replaced, for instance, by a consistent estimate of $\phi$,
such the maximum likelihood estimate $\hat \phi$.

As a residual proposal we can work, for instance, with the deviance component residual similarly to the univariate
case. Svetliza and Paula (2003) performed various simulation studies with indication of a very good
agreement between the empirical distribution of the deviance component residual and the normal distribution,
even for $\phi$ small. In the multivariate case we will adopt the following expression for the
deviance component residual:
$$
d^*(y_{ij}; \hat \mu_{ij}, \hat \phi) =  \frac{d(y_{ij}; \hat \mu_{ij}, \hat \phi)}{\sqrt{1 - \hat h_{ijj}}},
$$
where
$$d(y_{ij}, \hat \mu_{ij}, \hat \phi) = \pm\sqrt{2}\{d^2(y_{ij}, \hat \mu_{ij}, \phi)\}^{1/2},$$
in that $d^2(y_{ij}, \hat \mu_{ij}, \phi)$ is defined in (\ref{eq9}), $h_{ijj}$ is the $j$th principal diagonal element of the
matrix $\Hn_i = \Wn_i^{1/2} \Xn_i (\Xn_i^{\top} \Wn_i \Xn_i)^{-1} \Xn_i^{\top} \Wn_i^{1/2}$ and the sign is
the same of $(y_{ij} - \hat \mu_{ij})$.

A suggestion in order to assess departures from the postulated distributions for the responses $y_{ij}'s$, as well as
the presence of outlying observations, is to perform the normal probability plot for $d(y_{ij}, \hat \mu_{ij}, \hat \phi)$
with generate envelope (see, for instance, Svetliza and Paula, 2003). Another possibility, suggested by Waller and Zelterman (1997), is the Pearson residual  whose expression is given by
$$
   r_{ij} = \frac{(y_{ij} - \hat \mu_{ij})}{\sqrt{ \hat \mu_{ij} + \hat \phi^{-1} \hat \mu_{ij}^2}}
$$
Again, here it is recommended to perform the normal probability plot with generated envelope to detect
possible departures from the error assumptions as we as outlying observations.

\section{Applications}
\subsection{Epilepsy data}
Diggle et al. (2002) described an experiment in which 59 epileptic patients were randomly assigned to one of two
treatment groups: treatment (progabide drug) and placebo groups.
The number of seizures experimented by each patient during the baseline period (eight-week) and the four consecutive periods (two-week) was recorded. The main objective of this application is to analyze the drug effect and compare its effect with the placebo group effect. Overdispersion evidences were observed in the data set and a generalized estimating equation model was
applied to fit the data. In order to illustrate the potentiality of the random intercept Poisson-GLG model we modify the data set,
patient \#49 was dropped and some values of the patient \#18 were modified in order to make it an outlying observation.
Then, similarly to Diggle et al. (2002), we assume the following random intercept
Poisson-normal model:
\begin{figure}[h]
\centerline{\includegraphics[scale=0.5]{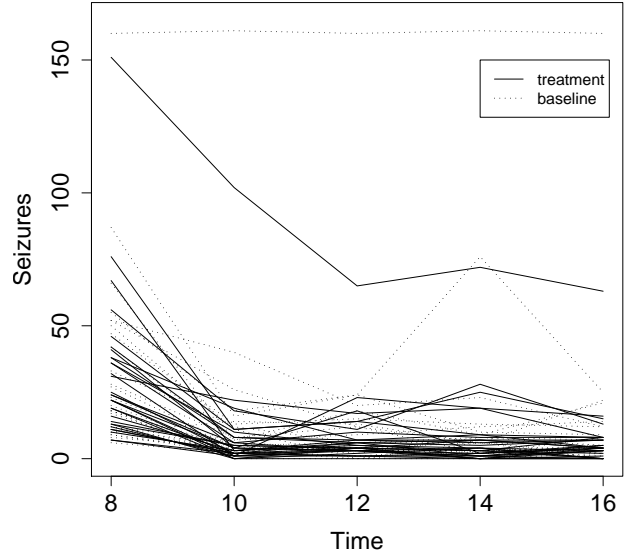}}
\caption{Profile of the seizures of the placebo and treatments groups.}
\end{figure}
\begin{description}
\item (i) $y_{ikj} | { b_i}$  $\stackrel{\rm ind} {\sim}$ ${\rm P}(u_{ikj})$,
\item (ii) $u_{i10} = {\rm exp}(\alpha + {b_i}) + {\rm log}(t_0), $\\
           $u_{i1j} = {\rm exp}(\alpha + \beta + {b_i}) + {\rm log}(t_j),$  $i = 1, \ldots, 28,$\\
           $u_{i20} = {\rm exp}(\alpha + { b_i}) + {\rm log}(t_0), $\\
           $u_{i2j} = {\rm exp}(\alpha + \beta + \delta + { b_i}) + {\rm log}(t_j)$  $i = 29,\ldots, 59,$ and
\item (iii) ${ b_i} \stackrel{\rm iid} {\sim} {\rm N}(0, \sigma^2),$
\end{description}
in that, $y_{ikj}$ denotes the number of seizures experienced by the $i$th patient in the $k$th group and
$j$th period, where $i = 1,\ldots,59,$ $k = 1, 2$ (placebo(28) and treatment(31)) and $j = 1, 2, 3, 4$.
In addition, $t_j$ denotes the week number
of the $j$th period ($t_0=8$ and $t_1 = t_2=t_3=t_4=2$), $\beta$ is the parameter which
represents the treatment effect and $\delta$ is the parameter referents to the
treatment group effect in relation to the placebo group.

However, from Figure 2, that describes
the log(counting) of seizures in the baseline period, we notice a skew form to right suggesting a skew
distribution for the random intercept. Thus, a random intercept Poisson-CLG model (with $\lambda < 0$) is also
assumed to fit this data set. Indeed, we replace in the model (i)-(iii)
the assumption ${b_i} \stackrel{\rm iid} {\sim} {\rm N}(0, \sigma^2)$ by
${b_i} \stackrel{\rm iid} {\sim} {\rm GLG}(0, \sigma, \lambda)$ with $\lambda < 0$.
\begin{figure}[h]
\centerline{\includegraphics[scale=0.5]{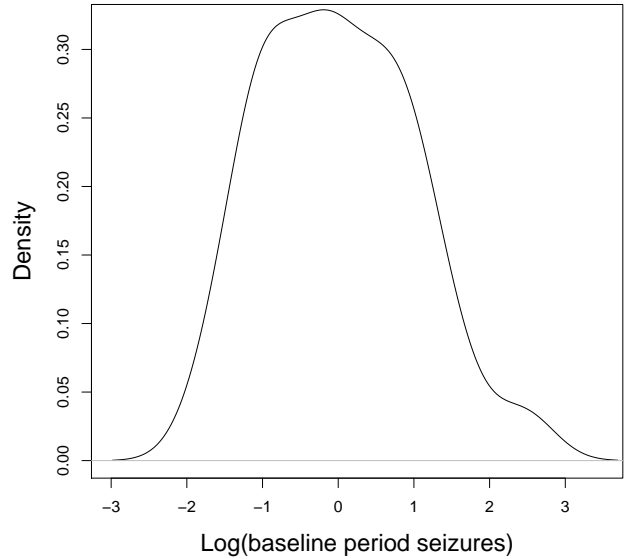}}
\caption{Density of the log(counting) of seizures in the baseline period for the modified data (right).}
\end{figure}
The parameter estimates, which were obtained by using the procedure NLMIXED in SAS, are described in Table 2.
Even though the inferential conclusions are the same for both models, indicating a significant effect for the treatment group
in the sense of decreasing the seizure mean, the random intercept Poisson-GLG model seems to fit better the data under
the AIC criterion. Furthermore, the estimate $\hat{\lambda} = -1.1852$, that is significant at 5\%,
confirms the evidences of Figure 2 on a skew form to right for the random intercept distribution.

\begin{center}
\begin{tabular}{crcrrrrcr}\\
\multicolumn{9}{c}{\bf Table 2}\\
\multicolumn{9}{c}{Parameter estimates with the respective approximate standard errors for the  Poisson-normal}\\
\multicolumn{9}{c}{ and Poisson-GLG random intercept models fitted to modified epilepsy data.}\\ \hline
  &  &  Poisson-normal  &        &&&  & Poisson-GLG  & \\ \hline
Parameter   & Estimate    &Sd. error   & z-value  &&&Estimate  & Sd. error   & z-value  \\\hline
$\alpha$    & 0.9379      &0.1132       &8.29    &&& 0.5062     &  0.2049      &   2.47    \\
$\beta$     & 0.4841      &0.0418       & 11.59  &&& 0.4812     &  0.0416      &  11.56       \\
$\delta$    &-0.4741      &0.0610       &-7.78   &&&-0.4685     &  0.0606      &  -7.72     \\
$\sigma$    &0.8426       &0.0802      & -   &&& 0.6130     &  0.1418      &  -   \\
$\lambda$   &             &             &        &&&-1.1852     &  0.6024      &  -1.97     \\
  &  & AIC = 2214.62&                &  &&& AIC = 2155.04&      \\  \hline
\end{tabular}
\end{center}

\subsection{C. dubia data}
As a second illustration we will consider the data set described in Lange et al. (1994)
(see also See and Bailer, 1988),
which was obtained from a reproductive aquatic toxicology experiment and
whose aim was to study the effect of the herbicide Nitrofen on
the asexual reproduction of the freshwater invertebrate {\em Ceriodaphnia dubia} ({\em C. dubia}).
The data represent the offspring born counting in three broods to each of 10 {\em C. dubia} in each of 5
concentration groups of Nitrofen: 0, 80, 160, 235 and 310 $\mu g/l$.
The eggs of these species were developed and hatched within 48 hours
and were released from the brood pouch according to the molting cycle of adult female (24 to 48 hours).
The mean profiles of the offspring born counting under each concentration across the time are presented in Figure 3.

Various Poisson regression models with log link were applied by Lange et al. (1994)
in order to compare the Nitrofen
level potencies and the toxin effects on the individual organisms. A Poisson regression model with
a quadratic effect for the concentration was considered as the best-fitting model, but
the possibility of within-animal correlation (one has three broods for each animal) was not considered.
The normal probability plot for the deviance residual with generated envelope in Figure 4, indicates that the Poisson model
is not suitable to fit this data set, with indication of overdispersion.
Thus, the multivariate negative binomial (MNB) model appears as an option
to explain the offspring counting and based on the behavior of Figure 3 we suggest the
following model to fit the {\em C. dubia} data:
\begin{description}
\item (i) $\yn_i \stackrel{\rm ind} {\sim} {\rm MNB}(\mun_i, \phi)$ with
\item (ii) $\mu_{ij}$ = exp$(\beta_0$ + $\beta_1{\rm C}_{ij}$ + $\beta_2{\rm Day}_{ij}$
+ $\beta_3{\rm C}_{ij}\times {\rm Day}_{ij}),$
\end{description}
where $y_{ij}$ is the offspring counting of the
$i$th adult female in the $j$th brood, for $i = 1,\ldots,50$ and $j=1,2,3,$ with
C$_{ij}$ denoting the concentration for which the $j$th brood of the $i$th animal was submitted.
One has the following settings: C$_{ij} =0$ $(i=1, \ldots, 10)$,
C$_{ij} = 80$ $(i=11, \ldots, 20)$, C$_{ij}=160$ $(i=21, \ldots, 30)$,
C$_{ij}=235$ $(i=31, \ldots, 40)$ and
C$_{ij}=310$ $(i=41, \ldots, 50)$, for $j=1,2,3$,
and Day$_{ij}$
denotes the day in which the eggs were hatched for the $j$th brood of the $i$th animal and
it assumes the following values:
Day$_{i1}=-2$, Day$_{i2}=0$ and Day$_{i3}=2$, for $i=1, \ldots, 50$.
\begin{figure}[h]
\centerline{\includegraphics[scale=0.5]{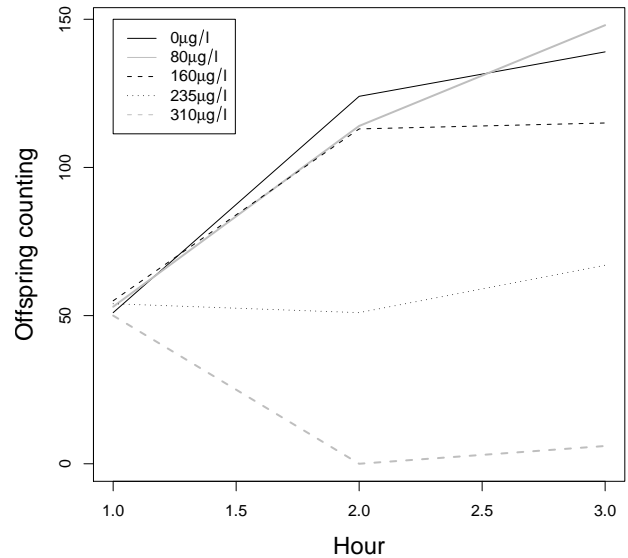}}
\caption{Mean profiles of the offspring born counting of {\em C. dubia}
under each concentration across the time.}
\end{figure}
\begin{figure}[h]
\centerline{\includegraphics[scale=0.5]{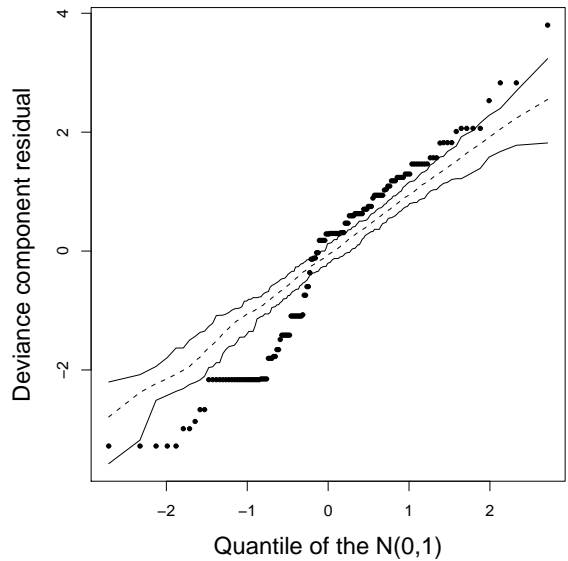}}
\caption{Normal probability plot with generated envelope for the Poisson quadratic
regression model fitted to {\em C. dubia} data.}
\end{figure}

\begin{figure}[!htb]
\begin{minipage}[b]{0.5\linewidth}
\includegraphics[width=\linewidth]{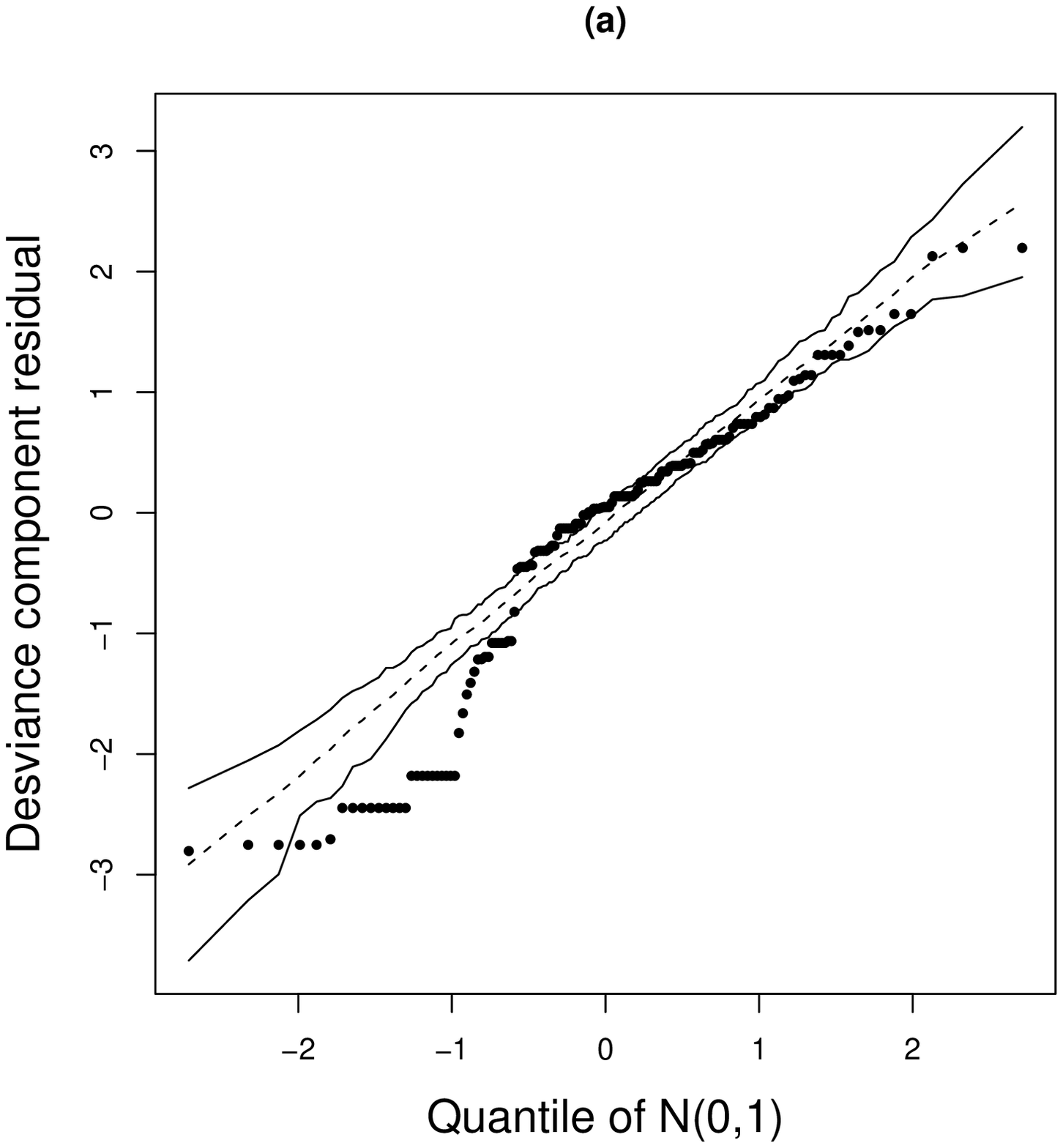}
\end{minipage} \hfill
\begin{minipage}[b]{0.5\linewidth}
\includegraphics[width=\linewidth]{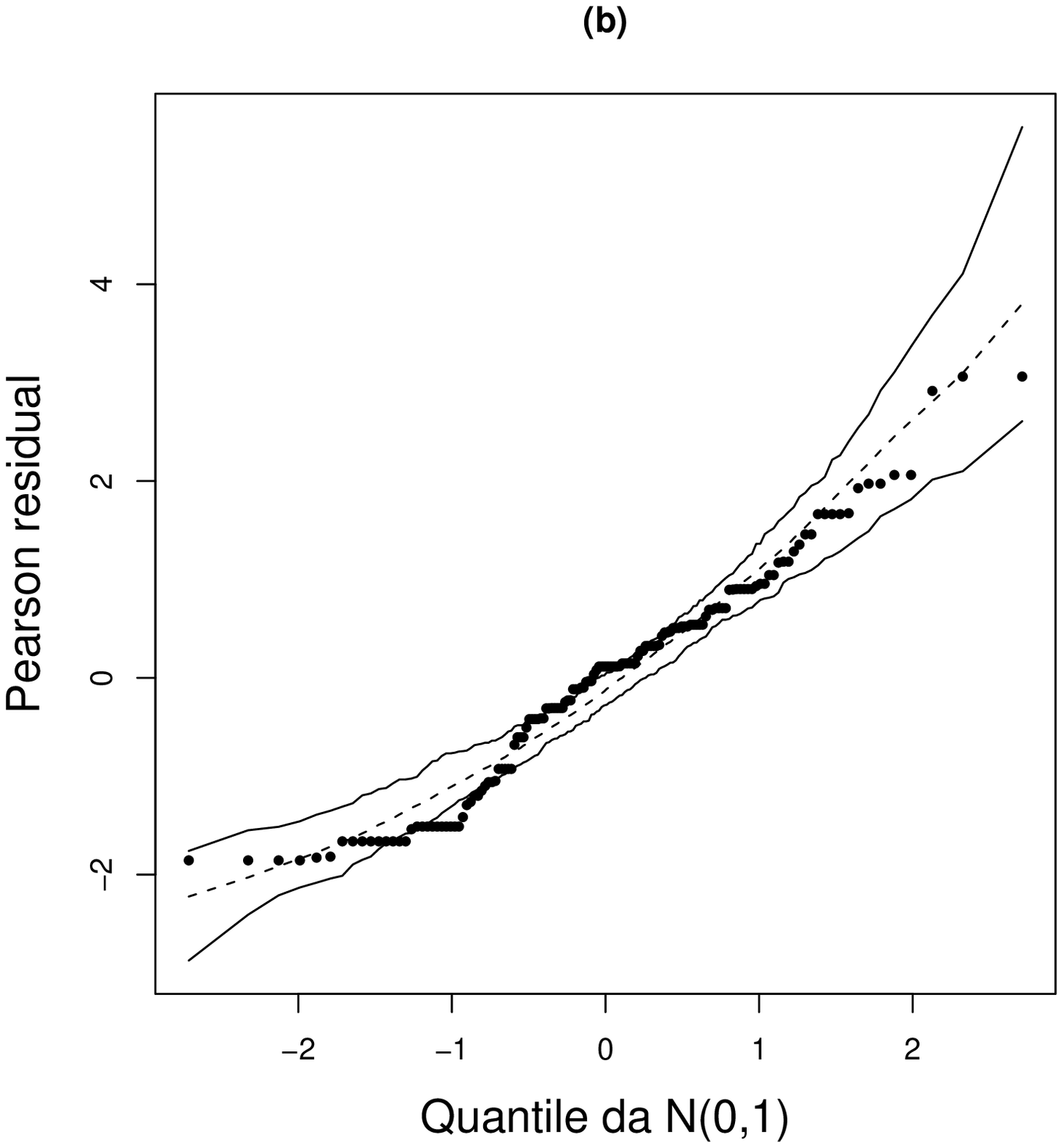}
\end{minipage}
\caption{Normal probability plot with generated envelope for
the univariate negative binomial model (left) and
multivariate negative binomial model model (right) fitted to {\em C. dubia} data. \label{fig2}}
\end{figure}

\begin{center}
\begin{tabular}{crcrrrrcr}\\
\multicolumn{9}{c}{\bf Table 3}\\
\multicolumn{9}{c}{Parameter estimates with the respective approximate standard errors for the negative binomial}\\
\multicolumn{9}{c}{ (NB) and multivariate negative binomial (MNB) models fitted to {\em C. dubia} data.}\\ \hline
  &  &  NB model  &        &&&  &  MNB model  & \\ \hline
Parameter   & Estimate  &Sd. Error   & z-value  &&&Estimate  & Sd. Error   & z-value  \\\hline
$\beta_0$   &2.5269    &0.0704      &35.8950    &&& 2.5333  &0.0925      &27.3619     \\
$\beta_1$   &-0.0040   &0.0004      &-9.8020    &&& -0.0040 &0.0005      &-7.5045      \\
$\beta_2$   &0.3329    &0.0432      &7.7050     &&& 0.2916  &0.0297      &9.8088    \\
$\beta_3$   &-0.0015   &0.0003     &-6.2020    &&&-0.0013  &0.0002      &-6.8379   \\
$\phi$      & 7.4116   & 2.1900     & -     &&&11.5603  &4.2267      & -  \\
AIC         &          & 830.3             &     &&&         & 829                  &\\
Deviance    &           &  225.76 (146 d.f.)&     &&&         & 222.91 (146 d.f.)     &\\\hline
\end{tabular}
\end{center}
The parameter estimates (standard errors) of the NB and MNB models, given in Table 3,
are similar, confirming the tendencies observed in Figure 3.
However, the AIC and deviance values suggest that the MNB model fitts better the data.
This can also be observed by comparing the normal probability plots for the deviance component and Pearson residuals
in Figure 5. Lange et al. affirms that the interclass correlations in the C. dubia data is small. For $\hat{\phi} = 11.5603,$ we identify moderate to weak intraclass correlations with the increase of the concentration levels. We also obtain $\hat{\lambda} = 0.293$$(0.053),$ a since that $\hat{\phi} = \hat{\lambda}^{-2}.$ When $\hat{\lambda}$ assumes a small value the MNB model has the feature of fits data sets positively correlated with a considerable number of zeros.
The data set C.dubia contains many zeros in the highest concentration level, however, some of the intraclass correlations are nearly zero. This fact can explain the lack of fit observed in Figure 5. The deviance component residual has been suggested in this paper was not used because $d^2(y_{ij}, \hat{\mu}_{ij}, \hat{\phi})$ assumed any negative values. We are investigating this fact.

\section{Concluding remarks}
In this paper we propose the generalized log-gamma distribution to
give flexibility for the random intercept
distribution in Poisson mixed models.
The advantage of this distribution is the skew forms to the right and to the left including the normal
distribution as a particular case. From the random intercept Poisson-GLG  model the multivariate negative binomial model
was derived for a particular parameter setting and the score functions, Fisher information matrix as well as an iterative process were
derived. Residual analysis were also proposed.
In addition, we present two motivating examples emphasizing the specials features of each model.
Particularly, for the epilepsy application, we conclude that the random intercept Poisson-GLG
model seems to be more appropriate to fit the modified data with indication of skew form for the intercept distribution as well as
presence of outlying observation.
In the {\em C. dubia} application the multivariate negative binomial model
seems to fit better the data  than the univariate negative binomial model,
since it incorporates the intraclass correlation that is in general  positive.
Thus, we believe that the models proposed in this work enlarge the options in the class of generalized linear mixed models
particularly to fit count data with indication of overdispersion and nonnormal distribution for the random effects.
Extensions for other responses, such binomial and gamma, as well as for two or more random effects are in progress.
\vspace{5mm}

\noindent {\tt Acknowledgment}: This work was supported by
CNPq and FAPESP, Brazil.

\section*{References}
\begin{description}
\item Ahn, H., 1996. Log-gamma regression modeling through regression trees.
Communications in Statistics, Theory and Methods 25, 295-311.

\item Alonso, A., Liti\`ere, S., Molenbergs, G., 2008. A family of tests to detect misspecification
in the random-effects structure of generalized linear mixed models.
Computational Statistics and Data Analysis 52, 4487-4501.

\item Breslow, N. E. and Clayton, D. G., 1993. Approximate inference in generalized linear mixed models.
Journal of the American Statistical Association 88, 9-25.

\item Chien-Tai, L., Wu, S. J. S., Balakrishnan, N., (2004). Interval estimation of
parameters of log-gamma distribution based on progressively censored data.
Communications in Statistics, Theory and Methods 33, 2595-2626.

\item Cox, C., Chu, H., Shneider, M. F., Munoz, A., 2007. Parametric survival analysis and
taxonomy of hazard functions for the generalized gamma distributions.
Statistics in Medicine 26, 4352-4374.

\item DiCiccio, T. J., 1987. Approximate inference for the generalized gamma distribution.
Technometrics 29, 33-40.

\item Diggle, P., Heagerty P., Liang, K. Y., Zeger, S., 2002. Analysis of Longitudinal Data.
Oxford Statistical Science Series, New York.

\item Johnson, N. L., Kotz, S., Balakrishnan, N., 1997. Discrete Multivariate Distributions.
Wiley, New York.

\item Lange, N., Ryan, L., Billard, L., Brillinger, D., Conquest, L., Greenhouse, J., 1994.
Case Studies in Biometry. Wiley, New York.

\item Lawless, J. F., 1980. Inference in the generalized gamma and log-gamma distributions.
  Technometrics 22, 409-419.

\item Lawless, J. F., 1987. Negative binomial and mixed Poisson regression.
Canadian Journal of Statistics 15, 209-225.

\item Lawless, J. F., 2002. Statistical Models and Methods for Lifetime Data, 2nd Edition.
Wiley, New York.

\item Lee, Y., Nelder, J. A., 1996. Hierarchical generalized linear models.
 Journal of the Royal Statistical Society B 58, 619-678.

\item Lee, Y., Nelder, J. A., 2001. Hierarchical generalized linear models : a synthesis of generalized
linear models, random effect models and structured dispersions.
Biometrika 88, 987-1006.

\item Litière, S., Alonso, A., Molenbergs, G., 2008. The impact of a misspecified random-effects
distribution on the estimation and the performance of inferential procedures in generalized linear mixed models.
Statistics in  Medicine 27, 3125-3144.

\item McCullagh, P., Nelder, J. A. (1989). Generalized
Linear Models, 2nd. Edition. Chapman and Hall, London.

\item McCulloch, C. E., Searle, S. R., 2001. Generalized Linear, and Mixed Models.
Wiley, New York.

\item Molenberghs, G., Verbeke, G., Demétrio, C. G. B., 2007. An extended random-effects approach to modeling repeated,
overdispersed count data. Lifetime Data Analysis 13, 513-531.

\item Ortega, E. M. M., Bolfarine, H., Paula, G. A., 2003. Influence diagnostics in
generalized log-gamma regression models.
Computational Statistics and Data Analysis 42, 165-186.

\item Ortega, E. M. M., Cancho, V. G., Paula, G. A., 2009. Generalized log-gamma
regression models with cure fraction.
Lifetime Data Analysis 15, 79-106.

\item Prentice, R., 1974. A log gamma model and its maximum likelihood estimation.
Biometrika 61, 539-544.

\item See, K., Bailer, A. J., 1988. Added risk and inverse estimation for count responses in reproductive aquatic
toxicology studies. Biometrics 54, 67-73.

\item Svetliza, C. F., Paula, G. A. (2003). Diagnostics in nonlinear negative
binomial models. Communications in Statistics,  Theory and Methods 32, 1227-1250.

\item Young, D. H., Bakir, S. T., 1987. Bias correction for a generalized
log-gamma regression model. Technometrics 29, 183-191.

\item Waller, A. L., Zelterman, D., 1997. Log-linear modeling with the negative multinomial
distribution. Biometrics 53, 971-982.

\item Zhang, P., Song, P. X. K., Qu, A., Greene, T., 2008.
Efficient estimation for patient-specific rates of disease progression
using nonnormal linear mixed models. Biometrics 64, 29-38.

\end{description}

\end{document}